\documentclass[conference,a4paper]{IEEEtran}

\IEEEoverridecommandlockouts

\usepackage[utf8]{inputenc}

\usepackage{amsmath,amssymb,epsfig,psfrag,cite,subfigure}
\include{macros}
\usepackage{graphicx}
\usepackage{epstopdf}
\usepackage{amsfonts}
\usepackage{multirow}
\usepackage[compact]{titlesec}
\usepackage[normalem]{ulem}
\usepackage{pgfplots}
\usepackage{acronym}

\usepackage{bm}

\usepackage{geometry}
\usepackage{tcolorbox}

\usepackage{accents}

\usepackage{hyperref}

\allowdisplaybreaks

\usepackage{soul}

\usepackage{amsthm}

\usepackage{nicefrac}

\usepackage{lipsum,multicol}

\usepackage{algorithm,balance}
\usepackage{algpseudocode}

\algdef{SE}[SUBALG]{Indent}{EndIndent}{}{\algorithmicend\ }%
\algtext*{Indent}
\algtext*{EndIndent}

\usepackage{enumitem}
\usepackage{mwe}    
\usepackage{epsfig,psfrag}
\usepackage{subfigure}
\usepackage{color}
\usepackage{url}
\usepackage{mathtools,xparse}

\usepackage{gensymb}

\usepackage[multiple]{footmisc}

\usepackage{xr}

\makeatletter
\newcommand*{\addFileDependency}[1]{
  \typeout{(#1)}
  \@addtofilelist{#1}
  \IfFileExists{#1}{}{\typeout{No file #1.}}
}
\makeatother

\newcommand*{\myexternaldocument}[1]{%
    \externaldocument{#1}%
    \addFileDependency{#1.tex}%
    \addFileDependency{#1.aux}%
}

\myexternaldocument{supp}

\usepackage{xcolor}

\newcommand\norm[1]{\left\lVert #1\right\rVert}

\newcommand{\hh}{\boldsymbol{h}}

\newcommand{\dd}{\boldsymbol{d}}
\newcommand{\yy}{\boldsymbol{y}}
\newcommand{\xx}{\boldsymbol{x}}
\newcommand{\zz}{\boldsymbol{z}}

\newcommand{\ww}{\boldsymbol{w}}

\newcommand{\BB}{\boldsymbol{B}}

\newcommand{\QQ}{\boldsymbol{Q}}

\newcommand{\bb}{\boldsymbol{b}}

\newcommand{\etab}{\boldsymbol{\eta}}

\newcommand{\deltaf}{\Delta_f}

\newcommand{\Ts}{T_s}

\newcommand{\complexset}[2]{ \mathbb{C}^{#1 \times #2}  }

\newcommand{\realp}[1]{ \Re \left\{#1\right\}  }

\newcommand{\trp}{\mathsf{T}}

\newcommand{\mtcn}{{\mathcal{CN}}}

\newcommand{\AAb}{\boldsymbol{A}}
\newcommand{\1}{\boldsymbol{1}}
\newcommand{\0}{\boldsymbol{0}}

\newcommand{\RR}{\boldsymbol{R}}

\makeatletter \renewcommand\d[1]{\ensuremath{%
		\;\mathrm{d}#1\@ifnextchar\d{\!}{}}}
\makeatother

\makeatletter
\newcommand*\rel@kern[1]{\kern#1\dimexpr\macc@kerna}
\newcommand*\widebar[1]{%
  \begingroup
  \def\mathaccent##1##2{%
    \rel@kern{0.8}%
    \overline{\rel@kern{-0.8}\macc@nucleus\rel@kern{0.2}}%
    \rel@kern{-0.2}%
  }%
  \macc@depth\@ne
  \let\math@bgroup\@empty \let\math@egroup\macc@set@skewchar
  \mathsurround\z@ \frozen@everymath{\mathgroup\macc@group\relax}%
  \macc@set@skewchar\relax
  \let\mathaccentV\macc@nested@a
  \macc@nested@a\relax111{#1}%
  \endgroup
}
\makeatother

\theoremstyle{remark}

\newtheoremstyle{mytheoremstyle} 
    {\topsep}                    
    {\topsep}                    
    {\upshape}                   
    {.5em}                           
    {\itshape}                   
    {.}                          
    {.5em}                       
    {}  
    
\theoremstyle{mytheoremstyle}

\newtheoremstyle{iremark}
  {\topsep}   
  {\topsep}   
  {\upshape}  
  {0.2in}       
  {\itshape}  
  {.}         
  {5pt plus 1pt minus 1pt} 
  {\thmname{#1}\thmnumber{ \itshape#2}\thmnote{ (#3)}} 

\theoremstyle{iremark}

\acrodef{RIS}{reconfigurable intelligent surface}
\acrodef{SNR}{signal-to-noise ratio}
\acrodef{ISAC}{integrated sensing and communications}
\acrodef{SLAM}{simultaneous localization and mapping}
\acrodef{ISLAC}{integrated sensing, localization, and communications}
\acrodef{LoS}{line-of-sight}
\acrodef{NLoS}{non-line-of-sight}
\acrodef{AoA}{angle-of-arrival}
\acrodef{AoD}{angle-of-departure}
\acrodef{UE}{user equipment}
\acrodef{NF}{near-field}
\acrodef{BS}{base station}
\acrodef{MCRB}{misspecified Cram\'{e}r-Rao bound}
\acrodef{CRB}{Cram\'{e}r-Rao bound}
\acrodef{CRLB}{Cram\'{e}r-Rao lower bound}
\acrodef{LB}{lower bound}
\acrodef{ML}{maximum-likelihood}
\acrodef{MML}{mismatched maximum-likelihood}
\acrodef{DL}{downlink}
\acrodef{UL}{uplink}
\acrodef{MIMO}{multiple-input multiple-output}
\acrodef{MISO}{multiple-input single-output}
\acrodef{SISO}{single-input single-output}
\acrodef{SIP}{shift invariance property}
\acrodef{FIM}{Fisher information matrix}
\acrodef{RMSE}{root mean-squared error}
\acrodef{AWGN}{additive white Gaussian noise}
\acrodef{ADMM}{alternating direction method of multipliers}
\acrodef{LS}{least-squares}
\acrodef{SOC}{second-order cone}
\acrodef{CFO}{carrier frequency offset}
\acrodef{TX}{transmit}
\acrodef{RX}{receive}
\acrodef{CP}{cyclic prefix}
\acrodef{ISI}{intersymbol interference}
\acrodef{PSD}{power spectral density}
\acrodef{OFDM}{orthogonal frequency-division multiplexing}
\acrodef{OMP}{orthogonal matching pursuit}
\acrodef{DI}{Doppler ignorant}

\acrodef{HOSVD}{Higher-order singular value decomposition}
\acrodef{SVD}{singular value decomposition}
\acrodef{ESPRIT}{estimation of signal parameters via rotational invariance techniques}
\acrodef{DFBS}{dual-function base-station}
\acrodef{GLRT}{generalized likelihood ratio test}

\definecolor{purple}{RGB}{115, 0, 255}

\newcommand{\balpha}{\Bar{\alpha}}
\newcommand{\btau}{\Bar{\tau}}

\newcommand{\ziter}[1]{\zz^{(#1)}  }

\begin{document}
\bstctlcite{IEEEexample:BSTcontrol}

\title{Pilot Distortion Design for ToA Obfuscation in Uplink OFDM Communication}

\author{\IEEEauthorblockN{Mahmut Kemal Ercan\IEEEauthorrefmark{1}, Alireza Pourafzal\IEEEauthorrefmark{2}, Musa Furkan Keskin\IEEEauthorrefmark{2}, Sinan Gezici\IEEEauthorrefmark{1}, and Henk Wymeersch\IEEEauthorrefmark{2}}

\IEEEauthorblockA{\IEEEauthorrefmark{1} Department of Electrical and Electronics Engineering, Bilkent University, Turkey \\
\IEEEauthorrefmark{2}  Department of Electrical Engineering, Chalmers University of Technology, Sweden\\ 
Email: ercan@ee.bilkent.edu.tr\vspace{-0.5cm}
}
\thanks{This work is supported by Swedish Research Council (VR) under Grant 2024-04390 and 2023-03821, and Chalmers Transport Area of Advance, and the Scientific and Technological Research Council of Turkiye (TUBITAK) 1515 Frontier R\&D Laboritaries Support Program for Turk Telekom neXt Generation Technologies Lab (XGeNTT) under project number 5249902.}
}

\maketitle

\begin{abstract}
We study uplink orthogonal frequency-division multiplexing (OFDM) pilot distortion to deliberately obfuscate time-of-arrival (ToA) estimation at a single base station while preserving communication performance. We design a complex per-subcarrier distortion vector that increases sidelobes of the mismatched ambiguity function (MAF) relative to its mainlobe, using two objectives: the sidelobe-to-peak level ratio and the integrated sidelobe level. The design is subject to a transmit-power budget and a proximity (dissimilarity) constraint around the communication-optimal pilot. Communication impact is quantified by a capacity-motivated lower bound obtained from the linear minimum mean-squared error error covariance with a mismatched channel estimate. The resulting generalized fractional program is solved with Dinkelbach’s transform and a difference-of-convex update that yields a closed-form Karush–Kuhn–Tucker step. Simulations on a single-input single-output OFDM link show that the optimized distortions raise MAF sidelobes and degrade delay estimation, as validated by a mismatched maximum-likelihood ToA estimator, while incurring only marginal capacity loss over a broad signal-to-noise ratio range. The method requires no protocol changes or artificial path injection and provides a signal-level mechanism to control ToA observability under communication constraints.
\end{abstract}

\acresetall 
\section{Introduction}\label{sec_intro}
Privacy has long been a fundamental concern in communication systems \cite{gai2018privacy}. Most existing research has focused on data privacy, where an eavesdropper attempts to intercept and decode sensitive information transmitted between legitimate users. To mitigate such attacks, numerous methods have been proposed, including cryptographic techniques, data perturbation and artificial-noise injection, transmit scheduling, differential privacy mechanisms, and physical-layer security approaches such as power allocation and parameter encoding \cite{yan2025privacy, barnes2020fisher, gurgunoglu2021optimal, goken2018ecrb}. Beyond data confidentiality, recent advances in millimeter-wave (mmWave) and MIMO-OFDM technologies have enabled highly accurate localization of user equipment (UE) through channel estimation, which introduces a new form of privacy risk, location privacy, where a base station (BS) may infer or continuously track the position of a UE without consent \cite{checa2020location}. To counter this, two main groups of location privacy preservation techniques have been proposed: spoofing-based methods, which mislead the base station by generating false localization cues \cite{li2025delay}, and obfuscation-based methods, which deliberately distort localization-relevant parameters such as the angle of arrival (AoA), angle of departure (AoD), and time or time-difference of arrival (ToA/TDoA) \cite{italiano2025holotrace}.

\textit{Spoofing-based privacy preservation} aims to intentionally mislead the BS by forcing it to estimate an incorrect UE position. In such approaches, the UE modifies its transmission to create false but geometrically consistent channel observations. For instance, the fake-path injection technique \cite{li2024channel} and the delay–angle information spoofing method \cite{li2025delay,italiano2025holotrace} introduce artificial propagation paths or distort angle/delay features, misleading the BS into estimating false AoA/AoD or ToA/TDoA parameters. Similarly, selective-PRS spoofing \cite{gao2023your,gao2024surgical} exploits unauthenticated reference signals in 5G systems to manipulate ToA/TDoA estimates. Some other related works in physical-layer authentication \cite{pham2025leveraging, srinivasan2024aoa, pourafzal2025cooperative} also examined the feasibility of AoA spoofing under digital, analog, and hybrid array architectures. While spoofing can effectively deceive unauthorized BSs, it often requires precise channel state information (CSI) or additional complex procedures such as authorization. Furthermore, spoofing methods designed for the single BS cases can easily be circumvented in multiple BS systems. For example, additional delay shifts can be countered using TDoA \cite{ge2024experimental}. These limitations motivate the need for simpler, signal-level privacy mechanisms that preserve link reliability.
Alternatively, \textit{obfuscation-based privacy preservation} methods focus on degrading localization accuracy without producing specific false estimates. A notable example is the beamforming-based obfuscation strategy proposed in \cite{checa2020location}, which conceals AoDs to prevent direction estimation. In \cite{zhang2025privacy}, a Cramér-Rao-bound-based optimization framework was introduced to degrade localization performance of untrusted BSs while maintaining communication quality for legitimate ones. Experimental results with Wi-Fi systems \cite{ayyalasomayajula2023users} demonstrated that small phase or delay perturbations can reduce positioning precision with negligible throughput loss. Artificial-noise injection \cite{tomasin2022beamforming} was also studied but cannot achieve controlled location spoofing. Compared to these, spoofing-based techniques offer higher control over the perceived location but often at the cost of complexity and may not work for multiple BS environment.
Beyond privacy-preserving design,  optimization frameworks are also applied in integrated sensing and communication (ISAC) systems, where the transmit waveform is optimized to \textit{balance sensing and communication objectives} \cite{keskin2021limited,he2024dual}.

Despite the central role of ToA and TDoA estimation in 5G positioning frameworks, no existing method offers a unified strategy to deliberately degrade ToA estimation accuracy while maintaining reliable communication performance. In this work, we propose a location-privacy-preserving technique in which a single-antenna UE transmits pilot signals intentionally distorted to degrade ToA estimation at a single-antenna BS. Unlike conventional spoofing, our method does not inject additional delays but instead modifies the pilot waveform itself, increasing the side-lobe levels in the delay profile of BSs to increase ToA estimation error while maintaining reliable communication. 

\section{System Model and Problem Formulation}

We consider an uplink channel between a single-antenna UE and a single-antenna BS. This simplified setup excludes beamforming effects, as ToA estimation depends only on signal timing rather than spatial processing, allowing a clear focus on pilot distortion and its impact on delay estimation. The channel is modeled in a generic multipath form, accommodating multiple propagation paths with different gains and delays. The UE’s goal is to maintain acceptable communication quality while degrading the ToA estimation accuracy for privacy preservation, whereas the BS seeks both reliable communication and accurate ToA-based localization. 

\subsection{Transmit Signal Model with Distorted Pilots}
The uplink baseband transmit signal is modeled as an OFDM waveform, where the unit-modulus pilot symbols $x_i\in \mathbb{C},\; i=0,\hdots,N-1$, jointly known and agreed upon by the BS and the UE, are transmitted over $N$ subcarriers. To preserve location privacy, the UE secretly applies a complex distortion factor $z_i\in \mathbb{C}$ to each subcarrier, modifying the amplitude and/or phase of the pilot without the BS' knowledge. The overall transmit power is limited by the constraint $\sum_{i=0}^{N-1} |z_i|^2 \leq P_t,$ ensuring that the distorted pilots remain within the maximum power budget $P_t$ while enabling controlled degradation of localization performance. Data symbols are omitted, as they do not directly affect delay estimation or the localization metrics considered in this study. The resulting discrete-time OFDM pilot signal is as follows:
\begin{equation}
    s[k] = \frac{1}{N} \sum_{i=0}^{N-1}x_i z_i e^{j2\pi k i/N},\; k = 0,\hdots,N-1\;.
\end{equation}
The index $i$ corresponds to the subcarrier while $k$ corresponds to the sample index. Generally, a cyclic prefix is added to the beginning of each block, which means that the last $N_\text{cp}$ samples are repeated at the beginning. This extends $k = 0,\hdots,N-1$ to $k = -N_\text{cp},\hdots,-1,0,1,\hdots,N-1$. The transmit signal is converted to analog with $T$ rate, where $T$ is the duration of the elementary symbol. Accordingly, the continuous-time pilot signal is written as:
\begin{equation} \label{eq:carried_signal}
    s(t) =  \sum_{i=0}^{N-1}x_i z_i e^{j2\pi i\deltaf t},\; -T_\text{cp}\leq t \leq \Ts \,,
\end{equation}
where $\deltaf = 1/T$ being the subcarrier spacing, $T_\text{cp} = N_\text{cp}T$ and $\Ts = NT$ being the total OFDM symbol time.

\subsection{Received Signal Model}
The transmitted pilot block propagates through a multipath channel, where each path introduces an individual delay $\tau_l$ and attenuation $\alpha_l$, with a total number of $L$ paths resulting in the channel coefficients at the $i$th subcarrier as $h_i = \sum_{l=1}^L \alpha_l e^{-j 2\pi i\tau_l/T} $. After reception, the BS performs DFT-based demodulation to extract the subcarrier-domain signal. Accordingly, the received pilot signal in the frequency domain can be expressed as
\begin{align} \label{eq:rec_freq}
    \yy = (\xx\odot\zz)\odot\hh + \ww\,,
\end{align}
where the vectorized symbols $\yy, \xx, \zz, \hh, \ww\in\complexset{N}{1}$ are received signal, transmitted symbols, symbol distortion, channel, and noise, respectively. Here, $[\ww]_k = w_{k} \sim \mtcn(0, \sigma^2)$ is the complex additive Gaussian noise, where $\sigma^2 = N_0 N \deltaf$ being the noise variance with $N_0$ denoting the noise power spectral density (PSD) level. 

\subsection{Problem Formulation}
Given \eqref{eq:rec_freq}, our objective is to intentionally degrade the ToA estimation accuracy at the BS through the design of the distortion vector $\zz$ at the UE, while ensuring that the associated channel estimation error introduces only a negligible reduction in communication capacity, thereby preserving the reliability of the communication link. 
To analyze how pilot distortion impacts this estimation process, we consider a single BS link, which captures the fundamental ToA behavior without loss of generality. In a localization system, each receiving BS would be affected by the distorted pilots. 


\section{Performance Metrics}
\subsection{Communication Metric}

To evaluate the communication performance, we relate the achievable rate to the accuracy of the channel estimate obtained from the pilot signal. Since pilot distortion directly affects the channel estimation quality, it indirectly determines the achievable communication rate. To capture this dependency, we adopt the lower-bound formulation in \cite[Prop.~1]{keskin2021limited}, where the channel capacity is expressed as a function of the covariance matrix of the LMMSE estimate of a communication signal transmitted immediately after the pilot, experiencing the same channel conditions. The following describes the corresponding channel estimation and capacity formulation.

Using the observations and assumed pilot signals, we can estimate the channel \cite[Eq.~(3.52)]{barbarossa2005multiantenna}:
\begin{equation} \label{eq:hh_est}
    \hat{\hh} = \yy \oslash \xx \,,
\end{equation}
where $\oslash$ denotes the elementwise division\footnote{The estimation model here is slightly different from the estimation model in \cite{barbarossa2005multiantenna}. They study over the time domain sampled signals \cite[Eq.~(3.51)]{barbarossa2005multiantenna} and estimate the time domain channel response to perform zero-forcing (ZF) estimate of the channel. However, we use the frequency domain channel response over subcarriers. Therefore, we adapted \cite[Eq.~(3.24)]{barbarossa2005multiantenna}, which is ZF symbol estimate, for ZF channel estimate in our misspecified model.}.
Consider another OFDM symbol, having data symbols $\xx^\text{com}$. We can assume that the same channel coefficients $\hh$ of $\xx$ is applied to $\xx^\text{com}$ if they are transmitted sequentially. Now, we can find the lower bound of the capacity by using the covariance matrix of the LMMSE estimate of $\xx^\text{com}$ \cite[Prop.~1]{keskin2021limited}:
\begin{equation} \label{eq:capacity}
    C(\hh,\hat{\hh})\geq -\log\det \RR_\text{LMMSE}(\hh,\hat{\hh})\,,
\end{equation}
with the covariance matrix $\RR_\text{LMMSE}(\hh,\hat{\hh}) = \text{diag}(r_0,\ldots, r_{N-1})$ in which
\begin{align}
    r_i & = \frac{|\hat{h}_i|^2(|h_i|^2+\sigma^2)}{(|\hat{h}_i|^2+\sigma^2)^2} -\frac{2\realp{\hat{h}_i^*h_i}}{|\hat{h}_i|^2+\sigma^2} +1 \label{eq:Rlmmse1}\\
    & = \frac{|y_i/x_i|^2(|h_i|^2+\sigma^2)}{(|y_i/\tilde{x}_i|^2+\sigma^2)^2}-\frac{2\realp{(y_i/x_i)^*h_i}}{|y_i/x_i|^2+\sigma^2} +1 \label{eq:Rlmmse2}
\end{align}
where we used \eqref{eq:hh_est} to obtain \eqref{eq:Rlmmse2} and . Finally we find that  $ \det \RR_\text{LMMSE}(\hh,\hat{\hh}) = \prod_{i=0}^{N-1} r_i$. 
For our optimization, we focus on the estimation error induced solely by pilot distortion, and therefore neglect the explicit noise component in the estimated channel expression. Under this assumption, $\hat{\hh} \approx (\hh\odot\xx\odot\zz) \oslash \xx = \hh\odot\zz$ resulting in \cite[eq.~(20)]{keskin2021limited}
\begin{align}
    C(\hh,\hat{\hh})\geq 
&-\sum_{i=0}^{N-1} \log \psi_i(z_i,h_i) 
\end{align}
where 
\begin{align*}
    \psi_i(z_i,h_i)=\frac{|h_i|^2|z_i|^2(|h_i|^2+\sigma^2)}{(|h_i|^2|z_i|^2+\sigma^2)^2} -\frac{2|h_i|^2\realp{z_i}}{|h_i|^2|z_i|^2+\sigma^2} +1.
\end{align*}
Note that the channel capacity is affected by both amplitude and phases of elements of $\zz$. We  use this bound as a communication metric:
\begin{align}
    &f_\text{com}(\zz)=-\sum_{i=0}^{N-1} \log \psi_i(z_i,h_i).\label{eq:f_com} 
\end{align}

\subsection{ToA Estimation Metrics} \label{sec:loc_metrics}

The channel is modeled as a multipath channel comprising $L$ propagation components, each characterized by a complex gain and delay. To characterize the ToA estimation performance and the ability to resolve multipath, we note that 
%
delay estimation is typically performed by cross-correlation between the received and known pilot signals. Hence, the estimation accuracy can be characterized by the cross-ambiguity function between the distorted and assumed signals, which is itself closely related to the Cramér-Rao-bound \cite{sakhnini2024general}.
We refer to this as the \textit{mismatched ambiguity function} (MAF), as it accounts for potential mismatches between the true and assumed transmitted waveforms. The MAF can be defined as a function of delay $\tau$  as follows:
\begin{equation} \label{eq:maf}
    \chi (\tau) \triangleq \int_{-\infty}^\infty s(t) \tilde{s}(t-\tau) \text{d}t \,,
\end{equation}
where $s(t)$ is the transmitted signal given in \eqref{eq:carried_signal}, and $\tilde{s}(t)$ represents the assumed pilot model:
\begin{equation}
    \tilde{s}(t) = \sum_{i=0}^{N-1}x_ie^{j2\pi i\deltaf t}, \quad -T_\text{cp}\leq t\leq\Ts \,.
\end{equation}
After simplification, we obtain
\begin{equation}
    \chi (\tau) = \int_{-T_\text{cp}+\tau}^{\Ts} \sum_{k=0}^{N-1}\sum_{l=0}^{N-1} x_kx_l^*z_k e^{j2\pi\deltaf(kt-l(t-\tau))}\text{d}t \,.
\end{equation}
Neglecting constant scaling terms and performing simplifications 
yields the following compact form:
\begin{equation} \label{eq:maf_tau_simplified}
    \chi (\tau) = \sum_{k=0}^{N-1} z_k e^{j2\pi k\deltaf\tau} \,.
\end{equation}
%
The MAF in \eqref{eq:maf_tau_simplified} is a complex-valued function.
Since delay estimation depends on its magnitude response, we focus on the squared magnitude $|\chi (\tau)|^2$ which captures the correlation power at delay $\tau$.
To analyze its dependence on the distortion vector 
$\zz$, the MAF can be written as a linear function of $\zz$, i.e., $\chi(\tau) = \dd^{\mathsf{H}}(\tau) \zz$ which leads to
\begin{equation} \label{eq:af_quadratic}
    |\chi(\tau)|^2 = \zz^{\mathsf{H}} \QQ(\tau) \zz\,, 
\end{equation} 
where $\QQ(\tau) = \dd(\tau) \dd^{\mathsf{H}}(\tau) \succeq \0$. Hence, $|\chi(\tau)|^2$ is a quadratic function of $\zz$; and since $\QQ(\tau)$ is positive semidefinite, $|\chi(\tau)|^2$ is convex w.r.t.~$\zz$.

Several performance metrics can be derived from the MAF to characterize the ToA estimation behavior.
In the following, we introduce two such metrics based on the MAF and analyze their convexity properties and computational formulations.
\begin{itemize}
    \item \textit{Sidelobe-to-Peak Level Ratio (SLPR):}
A first useful metric to quantify the delay estimation performance is the SLPR, defined as the ratio between the highest sidelobe and the mainlobe of the MAF magnitude response:
\begin{equation}
    \text{SLPR} = \frac{\max_{\tau\in T_\text{SL}} |\chi (\tau)|^2}{|\chi (0)|^2} \,,
\end{equation}
where $T_\text{SL}$ denotes the sidelobe region.
Since $T_\text{SL}$ depends on the mainlobe region, the SLPR is not strictly convex with respect to the distortion vector $\zz$. However, in the considered formulation, the optimization is performed within a small neighborhood of the communication-optimal point, where the structure of the ambiguity function remains nearly unchanged. In this local region, the position of the dominant sidelobe, denoted by $\tau^*$, can be assumed to remain fixed. This assumption is also validated in simulation results (see Fig.~\ref{fig:rangeProf_slpr}), where the location of $\tau^*$ shows negligible variation for the considered distortion levels. Under this setting, the SLPR can be expressed in the generalized Rayleigh quotient form:
\begin{equation}
    \frac{\zz^{\mathsf{H}}\QQ(\tau^*)\zz}{\zz^{\mathsf{H}} \QQ(0)\zz}.
\end{equation}
where the sidelobe location $\tau^*$ can be determined by solving ${\partial|\chi (\tau)|^2}/{\partial\tau} = 0$ and selecting the solution corresponding to the second-highest peak of the MAF magnitude.

\item \textit{Integrated Sidelobe Level (ISL):}
Another useful performance metric is the ISL, which quantifies the total sidelobe energy relative to the mainlobe power:
\begin{equation}
    \text{ISL} = \frac{\int_{\tau\in T_\text{SL}} |\chi (\tau)|^2 \text{d}\tau}{|\chi (0)|^2}\,.
\end{equation}
To examine the convexity of the ISL metric, we recall from \eqref{eq:af_quadratic} that $|\chi (\tau)|^2$ can be written in quadratic form. Accordingly, the ISL can be expressed as
\begin{align}
    \text{ISL} &= \frac{\int_{\tau\in T_\text{SL}} |\chi (\tau)|^2 \text{d}\tau}{|\chi (0)|^2}
    =\frac{\int_{\tau\in T_\text{SL}} \zz^{\mathsf{H}}\QQ(\tau)\zz \text{d}\tau}{\zz^{\mathsf{H}}\QQ(0)\zz}\\
    &=\frac{\zz^{\mathsf{H}} \left(\int_{\tau\in T_\text{SL}} \QQ(\tau) \text{d}\tau\right) \zz}{\zz^{\mathsf{H}}\QQ(0)\zz}\,.
\end{align}
Since $\QQ(\tau)\succeq \0$, the numerator is convex for a fixed sidelobe region $T_\text{SL}$, and so is the denominator. Therefore, the ISL metric also takes the form of a generalized Rayleigh quotient, which enables optimization using standard fractional programming techniques.
\end{itemize}

\section{Pilot Distortion for Privacy Preservation and Communication Enhancement}

The joint communication–localization trade-off can be formulated in several ways, depending on which objective is prioritized. One approach is to maximize communication performance while constraining the localization accuracy to remain above a certain threshold; another is to maximize localization degradation subject to acceptable communication quality. In this work, we adopt the latter perspective: first, we determine the communication-optimal solution by maximizing $f_\text{com}(\zz)$, and then intentionally degrade the localization performance by maximizing a selected metric $f_\text{loc}(\zz)$, which in this study corresponds to the SLPR and ISL metrics.
Our goal is to design the distortion vector $\zz$ that simultaneously enhances communication efficiency and reduces ToA estimation accuracy, formulated as follows.

\subsection{Communication-optimized Solution}
First, we will find the optimal communication solution by solving the following problem:
\begin{align} 
    \max_{\zz}&\; f_\text{com}(\zz) \\
    \text{s.t.} &\; \sum_{i=0}^{N-1} |z_i|^2 \leq P_t 
\end{align}
where $f_\text{com}$ is defined in \eqref{eq:f_com}. Note that $f_\text{com}$ is non-convex and nonlinear in both amplitude and phase of elements of $\zz$. 
It can be easily verified that for $P_t\geq N$, the optimal solution corresponds to the no-distortion case, i.e., $\zz = \1$ \cite[Remark~2]{keskin2021limited}, as expected. 

Note that increasing $P_t$ does not yield a higher-SNR solution of the form $\zz = A\1$ with $A\propto \sqrt{P_t}$, as $\zz = A\1$ would lead to a proportional scaling of the channel estimate, which leads to an overly capacity estimate that cannot be supported by the real channel.  

\subsection{ToA degradation-optimized Solution}

After obtaining the communication-optimal solution, we aim to degrade the localization performance (here measured in terms of the ToA estimation performance) while keeping the distortion vector $\zz$ close to the communication-optimal point $\zz_\text{com}$. The objective is to increase the ToA estimation error, and thus reduce localization accuracy by maximizing a selected localization metric $f_\text{loc}(\zz)$. The corresponding optimization problem is formulated as
\begin{align} \label{eq:main_prob}
    \max_{\zz}&\; f_\text{loc}(\zz) \\
    \text{s.t.} &\; \norm{\zz}^2 \leq P_t \\
    &\; \norm{\zz - \zz_\text{com}}^2/ P_t \leq \epsilon 
\end{align}
where the second constraint enforces a bounded deviation from the communication-optimal point. We consider two localization metrics, SLPR and ISL, both expressible in the generalized Rayleigh quotient form:
\begin{equation} \label{eq_loc_metric}
    f_\text{loc}(\zz) = \frac{\zz^{\mathsf{H}}\AAb\zz }{\zz^{\mathsf{H}}\BB\zz}\,,
\end{equation}
where the matrices in the nominator and the denominator:
\begin{align} \label{eq:matrix_defs}
    \AAb &= \begin{cases}
        \QQ(\tau^*) \text{ if SLPR is used}\\
        \int_{\tau\in T_\text{SL}} \QQ(\tau) \text{d}\tau \text{ if ISL is used}
    \end{cases}\,, \\
    \BB &= \QQ(0) \,.
\end{align}
This definition results in the following formulation:
\begin{align} \label{eq:main_prob_rayleigh}
    \max_{\zz}&\; \frac{\zz^{\mathsf{H}}\AAb\zz }{\zz^{\mathsf{H}}\BB\zz}\\
    \text{s.t.} &\; \norm{\zz}^2 \leq P_t \\
    &\; \norm{\zz - \zz_\text{com}}^2/ P_t \leq \epsilon \,.
\end{align}
Although the SLPR and ISL metrics are neither convex nor concave, both take the generalized Rayleigh quotient form, allowing \eqref{eq:main_prob} to be efficiently solved using Dinkelbach’s transform.

\subsection{Localization solution with Dinkelbach's Transform} \label{sec:dinkelbach}
Fractional optimization problems of the form in \eqref{eq:main_prob_rayleigh} are commonly encountered in communication system design \cite{shen2018fractional}. Here, a similar approach is adopted to address our localization-oriented formulation. 

\subsubsection{Dinkelbach's Transform of \eqref{eq:main_prob_rayleigh}}
By applying Dinkelbach’s transform \cite{dinkelbach1967nonlinear}, the maximization problem in \eqref{eq:main_prob_rayleigh} can be rewritten as a weighted difference of the numerator and denominator terms:
\begin{equation} \label{eq:slpr_dintelbach}
\begin{split}
    \max_{\zz}&\; \zz^{\mathsf{H}}\AAb\zz - \beta\zz^{\mathsf{H}} \BB\zz\\
    \text{s.t.} &\; \norm{\zz}^2 \leq P_t \\
    &\; \norm{\zz - \zz_\text{com}}^2/ P_t \leq \epsilon 
\end{split}
\end{equation}
where $\beta$ is an auxiliary variable updated at each iteration as
\begin{equation}
    \beta^{(t+1)} \gets \frac{{\ziter{t}}^{\mathsf{H}}\AAb\ziter{t}}{{\ziter{t}}^{\mathsf{H}} \BB\ziter{t}}\,,
\end{equation}
with $\ziter{t}$ denoting the solution of \eqref{eq:slpr_dintelbach} at iteration $t$. The optimal solution of \eqref{eq:main_prob_rayleigh} is achieved when $\beta = {\zz^*\AAb\zz^*}/{{\zz^*}^{\mathsf{H}} \BB\zz^*}$ with $\zz^*$ being the optimal vector, thereby guaranteeing convergence of the iterative process to the solution for \eqref{eq:main_prob_rayleigh}. The solution for \eqref{eq:slpr_dintelbach} can be found by the Difference of Convex (DoC) Functions method.

\subsubsection{DoC-Based Solution to \eqref{eq:slpr_dintelbach}}
Since both $\AAb$ and $\BB$ are positive semidefinite and $\beta\geq 0, \forall \zz$, the objective in \eqref{eq:slpr_dintelbach} can be expressed as the difference of two convex functions:
\begin{align}
    g(\zz) &= \zz^{\mathsf{H}} \AAb \zz\,,\\
    h(\zz) &= \zz^{\mathsf{H}} \BB \zz\,.
\end{align}
Maximizing $g(\zz)- h(\zz)$ is equal to minimizing $h(\zz)- g(\zz)$ over the same feasible region.  

For $P_t> N$, the communication-optimal point satisfies $\zz_\text{com} = \1$, and the region defined by $\norm{\zz - \zz_\text{com}}^2/ P_t \leq \epsilon$ becomes a small subset of $\norm{\zz}^2 \leq P_t$ for small $\epsilon$. Thus, the single constraint $\norm{\zz - \1}^2/ P_t \leq \epsilon$ can be used.

At each iteration $t$, the DoC method linearizes $g$ around $\ziter{t}$, yielding the following update:
\begin{equation}
    \ziter{t+1}\gets\arg\min_{\zz \in \mathcal{Z} }\;
h(\zz)-\big[g(\ziter{t})+\langle\nabla g(\ziter{t}),\,\zz-\ziter{t}\rangle\big].
\end{equation}
with $\mathcal{Z} \triangleq \left\{ \zz \in \mathbb{C}^N \,\middle|\, \lVert \zz - \mathbf{1} \rVert^2 \leq \epsilon P_t \right\}.$ By dropping constant terms and substituting $\nabla g(\zz)=2\AAb\zz$, the iteration reduces to the following convex QCQP:
\begin{equation} \label{eq:DoC_simplified}
\min_{\norm{\zz-\1}^2\leq \epsilon P_t}\;
\beta \zz^{\mathsf{H}}\BB\zz - 2\realp{(\AAb\ziter{t})^{\mathsf{H}} \zz}.
\end{equation}
The closed-form solution can be derived via the KKT conditions, as outlined in Appendix~\ref{sec:doc_soln}.

\subsection{Mismatched Maximum-Likelihood (MML) Estimator}
To analyze whether the side-lobe level-based metrics in \eqref{eq_loc_metric} can effectively capture ToA estimation performance in practice, we have implemented the MML estimator to evaluate the accuracy of ToA estimation. The MML estimator can be written as
\begin{equation}
\begin{split}
    \hat{\etab}_\text{MML} (\yy) &\triangleq \arg\max_{\etab} \log \Tilde{p}(\yy|\etab)\,,\\
    &= \arg\min_{\etab} \norm{\yy-\hh(\etab)\odot\xx}^2\,.
\end{split}
\end{equation}
After some manipulations, the MML estimation of $\tau$ becomes
\begin{equation}
    \hat{\tau}_\text{MML} (\yy) = \arg\max_{\tau} \left|(\dd(\tau)\odot\xx)^{\mathsf{H}}\yy\right|^2 \,.
\end{equation}
Note that the MML estimator is equal to $|\chi (\tau)|^2$ when the noise components are discarded from the observation $\yy$.

\section{Simulation Results}

In this section, we evaluate the performance of the proposed communication and ToA estimation framework. 

\subsection{Scenario and Parameters}
Throughout the simulations, we consider a single-path scenario with $L=1$, i.e., only the line-of-sight (LoS) component is present. This setting enables a fair and interpretable comparison between the achievable communication capacity and the corresponding ToA root-mean-square error (RMSE). The LoS path is characterized by a delay and complex channel gain, given by $\btau = 500$m and $\balpha = \sqrt{\text{SNR}_\text{mag}}\,\sigma e^{j\pi/4}$ respectively, where $\text{SNR}_\text{mag}$ denotes the signal-to-noise ratio (SNR) converted from decibels to linear scale and a trivial phase shift of $\pi/4$ is added. The noise PSD level is $N_0= -174 \,\text{dBm/Hz}$ and the receiver noise figure is 8 dB, resulting in $\sigma=8.8\times 10^{-7}$. The total transmit power is constrained by $P_t = N$, which corresponds to the intuitive uniform allocation case yielding $\zz_\text{com} =\1$. The OFDM parameters are set as $N=64$, subcarrier spacing $\deltaf = 120$kHz, and symbol duration $T = 1/\deltaf$. Since we study the case $|x_i|^2=1,\forall i$, resulting in $f_\text{com}$ and $f_\text{loc}$ being independent of $\xx$, $\xx$ has no effect on our metrics.

\subsection{Results and Discussion}
\begin{figure}
    \centering
    \input{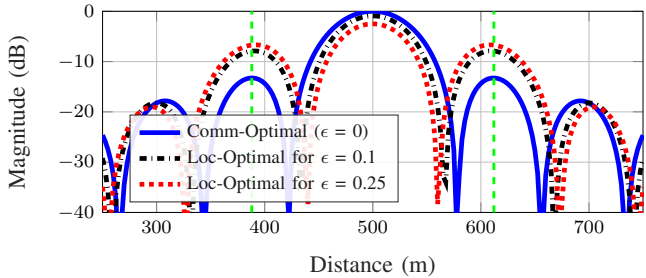}
    
    \caption{Range Profile for SLPR maximization with several $\epsilon$ values. Green lines denote $\tau^*$ in \eqref{eq:matrix_defs}.}
    \label{fig:rangeProf_slpr} 
\end{figure}

\begin{figure}
    \centering
    \input{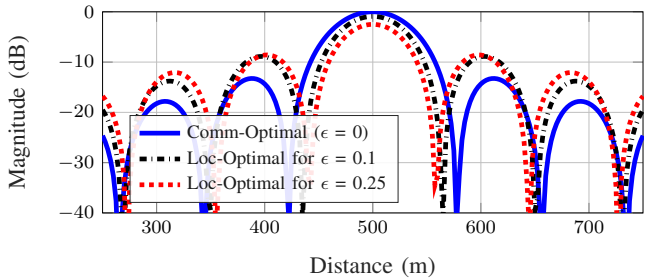}
    
    \caption{Range Profile for ISL maximization with several $\epsilon$ values.}
    \label{fig:rangeProf_isl} \vspace{-5mm}
\end{figure}

\noindent \textit{MAF Optimization Results:}
Using these parameters, we analyze the MAF for different optimal solutions of $\zz$. Figs. \ref{fig:rangeProf_slpr}-\ref{fig:rangeProf_isl} illustrate the range profiles corresponding to the communication-optimal and ToA estimation-degraded solutions obtained by optimizing the SLPR and ISL metrics, respectively, for two distinct values of $\epsilon$. Each figure depicts ${|\chi (\tau)|^2}/{N^2}$ evaluated over uniformly sampled delays $\tau\in [250/c , 750/c]$. 
From Fig.~\ref{fig:rangeProf_slpr}, it is observed that the assumption regarding the highest sidelobe point $\tau^*$ holds, with only a negligible shift. Furthermore, increasing $\epsilon$ leads to a higher sidelobe amplitude and a reduction in mainlobe level. In contrast, Fig.~\ref{fig:rangeProf_isl} demonstrates that ISL optimization elevates all sidelobe levels simultaneously. This behavior indicates that the proposed optimization framework effectively increases the MAF amplitude over the entire side-lobe region included in the definition of matrix $\AAb$ while decreasing the contribution associated with matrix $\BB$, which corresponds to the mainlobe point $\btau$ in this setup.

\begin{figure}
    \centering
%
%
\definecolor{mycolor1}{rgb}{1.00000,0.00000,1.00000}%
\definecolor{mycolor2}{rgb}{0.12941,0.12941,0.12941}%
\begin{tikzpicture}[scale=1\columnwidth/10cm,font=\footnotesize]
\begin{axis}[%
width=8 cm,
height=5 cm,
at={(0in,0in)},
scale only axis,
xmin=-5,
xmax=15,
xlabel style={font=\color{mycolor2}},
xlabel={SNR (dB)},
ymin=0,
ymax=3.5,
ylabel style={font=\color{mycolor2}},
ylabel={Capacity (Bits/s/Hz)},
axis background/.style={fill=white},
xmajorgrids,
ymajorgrids,
legend style={at={(0.01,0.55)}, anchor=south west, legend cell align=left, align=left, draw=white!15!black, fill opacity=0.9}
]
\addplot [color=blue, line width=2.0pt, mark size=5.0pt, mark=o, mark repeat=2, mark options={solid, blue}]
  table[row sep=crcr]{%
-5	0.274769892408345\\
-4	0.33511930078291\\
-3	0.406256284130706\\
-2	0.489167170311102\\
-1	0.584630709400607\\
0	0.693147180559945\\
1	0.814889218700011\\
2	0.949684188909912\\
3	1.09703181202892\\
4	1.25615333798053\\
5	1.42606243890537\\
6	1.6056450697986\\
7	1.79373623684702\\
8	1.98918491977485\\
9	2.19090267473672\\
10	2.39789527279837\\
11	2.6092793420315\\
12	2.82428726752588\\
13	3.04226384853904\\
14	3.26265782394902\\
15	3.48501071318057\\
};
\addlegendentry{$\text{Communication-Optimal (}\epsilon\text{ = 0)}$}

\addplot [color=red, dashed, line width=2.0pt, mark size=5.0pt, mark=triangle, mark repeat=2, mark options={solid, red}]
  table[row sep=crcr]{%
-5	0.257323427765189\\
-4	0.316044547544523\\
-3	0.386092817045489\\
-2	0.468761448794339\\
-1	0.565140381317084\\
0	0.675941241946536\\
1	0.801232350613468\\
2	0.940005088597445\\
3	1.0894505967533\\
4	1.24389122945731\\
5	1.39388875850816\\
6	1.52764988962672\\
7	1.63712517344972\\
8	1.724085204568\\
9	1.79730264367588\\
10	1.86523383585664\\
11	1.93284437284393\\
12	2.00210745838256\\
13	2.0732345374545\\
14	2.14557413084632\\
15	2.21813566919895\\
};
\addlegendentry{$\text{Loc-Optimal for SLPR, }\epsilon\text{ = 0.1}$}

\addplot [color=green, dashdotted, line width=2.0pt, mark size=5.0pt, mark=asterisk, mark repeat=2, mark options={solid, green}]
  table[row sep=crcr]{%
-5	0.21425679548757\\
-4	0.26313256623657\\
-3	0.32145603426189\\
-2	0.390346800096794\\
-1	0.470809546699654\\
0	0.563642439257735\\
1	0.669317274186097\\
2	0.787798341969878\\
3	0.918231318454123\\
4	1.05839738163081\\
5	1.20388019418823\\
6	1.34730194103014\\
7	1.47892960180998\\
8	1.58999002782644\\
9	1.67670274645704\\
10	1.74121345835123\\
11	1.78531715347564\\
12	1.8238058555494\\
13	1.84213905297138\\
14	1.84498045465941\\
15	1.86430182755382\\
};
\addlegendentry{$\text{Loc-Optimal for SLPR, }\epsilon\text{ = 0.25}$}

\addplot [color=orange, dashed, line width=2.0pt, mark size=5.0pt, mark=square, mark repeat=2, mark options={solid, orange}]
  table[row sep=crcr]{%
-5	0.262298830576383\\
-4	0.322450006073712\\
-3	0.394038160578886\\
-2	0.478169388140421\\
-1	0.575603599882558\\
0	0.686525479990102\\
1	0.810240931527676\\
2	0.944794001767537\\
3	1.08655630367435\\
4	1.22999785811589\\
5	1.36810496175316\\
6	1.49395942190079\\
7	1.6031650255627\\
8	1.69540855854416\\
9	1.77374292842344\\
10	1.84245147919267\\
11	1.90529574315777\\
12	1.96483230323415\\
13	2.02246769837516\\
14	2.07877844291224\\
15	2.13383391079985\\
};
\addlegendentry{$\text{Loc-Optimal for ISL, }\epsilon\text{ = 0.1}$}

\addplot [color=black, dashdotted, line width=2.0pt, mark size=5.0pt, mark=+, mark repeat=2, mark options={solid, black}]
  table[row sep=crcr]{%
-5	0.231665511584762\\
-4	0.286936774033713\\
-3	0.353571672990527\\
-2	0.433121965210701\\
-1	0.527059876413423\\
0	0.636609238194969\\
1	0.762428077987997\\
2	0.903977688129302\\
3	1.05825102055041\\
4	1.21735998805721\\
5	1.36493434450754\\
6	1.47448553470826\\
7	1.51901387136985\\
8	1.49379445465066\\
9	1.42347894936309\\
10	1.33797912723916\\
11	1.25495780331983\\
12	1.18161173624538\\
13	1.11992296528601\\
14	1.06964650424778\\
15	1.02965388937537\\
};
\addlegendentry{$\text{Loc-Optimal for ISL, }\epsilon\text{ = 0.25}$}

\end{axis}

\end{tikzpicture}%
    
    \caption{Channel capacity versus SNR for SLPR and ISL maximization with different $\epsilon$ values.}
    \label{fig:chancap_snr_all} \vspace{-5mm}
\end{figure}
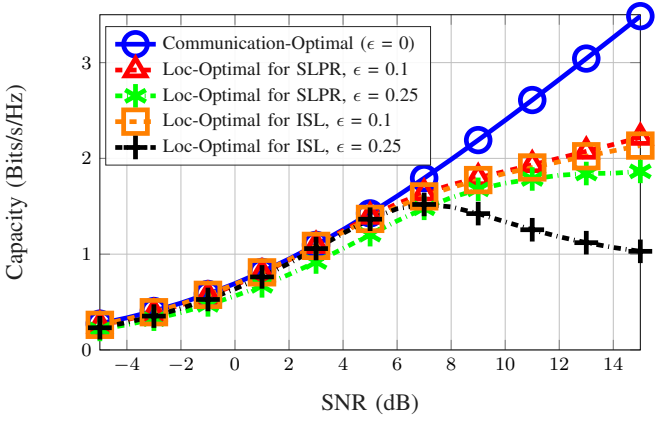

\noindent \textit{Trade-off Between Communication and ToA Estimation Performance:}
To assess the impact of optimizing $\zz$, we evaluate both communication and localization performance using the communication metric $f_\text{com}$ and the RMSE of the MML estimator. Figs.~\ref{fig:mml_snr_all} and \ref{fig:chancap_snr_all} present the RMSE and achievable capacity results for varying SNR values.
In Fig.~\ref{fig:mml_snr_all}, the RMSE is observed to increase with $\epsilon$, confirming that ToA estimation performance degrades as the distortion constraint is relaxed. A waterfall behavior is evident for all distortion levels, indicating that the estimator converges to the true delay value. This is expected, as the mainlobe remains dominant over the sidelobes for the selected $\epsilon$ values. However, when $\epsilon$ exceeds a certain threshold, the sidelobe magnitudes can surpass the mainlobe, leading to convergence to a higher RMSE value. Additionally, the SLPR-optimized solution with $\epsilon=0.25$ exhibits convergence to local sidelobe peaks at high SNRs, which is attributed to estimation outliers arising from elevated sidelobe levels. Since SLPR optimization produces higher sidelobes compared to ISL optimization (see Figs. \ref{fig:rangeProf_slpr} and \ref{fig:rangeProf_isl}), its overall RMSE is consequently higher.
A slight oscillatory trend can also be observed in Fig.~\ref{fig:mml_snr_all}, mainly because the optimized distortion vector $\zz$ changes abruptly even for small variations in $\epsilon$, which affects the overall MAF pattern and produces minor fluctuations in RMSE.

Fig.~\ref{fig:chancap_snr_all} illustrates that the achievable capacity is relatively robust in the studied SNR range, though a noticeable degradation occurs at high SNR values. It is also noted that for the ISL-optimized case with $\epsilon=0.25$, the capacity curve decreases slightly beyond approximately 7 dB. This behavior arises because the adopted capacity lower bound sums subcarrier-wise terms, some of which may become negative when $|z_i|$ is close to zero, and this effect becomes more pronounced at higher SNRs. Based on these results, the following guideline emerges: for high-SNR scenarios, smaller $\epsilon$ values should be preferred, as capacity is more sensitive to distortion in this regime; conversely, for low-SNR conditions, larger $\epsilon$ values are favorable, since ToA RMSE becomes more fragile than the communication capacity. This confirms that the proposed pilot distortion strategy can significantly degrade ToA accuracy while incurring only a negligible reduction in communication capacity.

\begin{figure}
    \centering
%
%
\definecolor{mycolor1}{rgb}{1.00000,0.00000,1.00000}%
\definecolor{mycolor2}{rgb}{0.12941,0.12941,0.12941}%
\begin{tikzpicture}[scale=1\columnwidth/10cm,font=\footnotesize]
\begin{axis}[%
width=8 cm,
height=5 cm,
at={(0.758in,0.481in)},
scale only axis,
xmin=-10,
xmax=5,
xlabel style={font=\color{mycolor2}},
xlabel={SNR (dB)},
ymode=log,
ymin=1,
ymax=1000,
yminorticks=true,
ylabel style={font=\color{mycolor2}},
ylabel={MML RMSE (m)},
axis background/.style={fill=white},
xmajorgrids,
ymajorgrids,
yminorgrids,
legend style={at={(0.37,0.55)}, anchor=south west, legend cell align=left, align=left, draw=white!15!black, fill opacity=0.9}
]
\addplot [color=blue, line width=2.0pt, mark size=5.0pt, mark=o,mark repeat=5, mark options={solid, blue}]
  table[row sep=crcr]{%
-10	483.19935806131\\
-9.5	430.866350057452\\
-9	379.847194032435\\
-8.5	324.275282181647\\
-8	264.406178587976\\
-7.5	194.986391854685\\
-7	163.706388518522\\
-6.5	108.129519071099\\
-6	53.4105501504389\\
-5.5	43.1424330351439\\
-5	9.59330123698341\\
-4.5	7.3231669824883\\
-4	6.27754004612317\\
-3.5	5.91297115199538\\
-3	5.57153099617619\\
-2.5	5.25132281679384\\
-2	4.95071226938166\\
-1.5	4.6682633454508\\
-1	4.40269655966035\\
-0.5	4.15285974911241\\
0	3.91770770472997\\
0.5	3.69628606424933\\
1	3.48771931729631\\
1.5	3.29120086682313\\
2	3.1059853916485\\
2.5	2.93138217448434\\
3	2.76674966305934\\
3.5	2.61149076766075\\
4	2.4650488076111\\
4.5	2.32690417423392\\
5	2.19657091636344\\
};
\addlegendentry{$\text{Communication-Optimal (}\epsilon\text{ = 0)}$}

\addplot [color=red, dashed, line width=2.0pt, mark size=5.0pt, mark=triangle,mark repeat=5, mark options={solid, red}]
  table[row sep=crcr]{%
-10	593.136145784088\\
-9.5	533.665406632448\\
-9	471.489068473208\\
-8.5	418.581154782094\\
-8	367.034403827707\\
-7.5	311.549288550564\\
-7	244.086256006705\\
-6.5	190.184730328507\\
-6	159.708061241093\\
-5.5	111.220026237912\\
-5	58.9327570723849\\
-4.5	43.7428824245268\\
-4	11.3113999685085\\
-3.5	8.19412935384802\\
-3	7.25702173961922\\
-2.5	6.11137870583802\\
-2	5.88794745107158\\
-1.5	5.68183561644592\\
-1	4.93943638964917\\
-0.5	3.81840882831567\\
0	3.60215218759361\\
0.5	3.39852982634459\\
1	3.20673392219347\\
1.5	3.02602114091287\\
2	2.85570480348511\\
2.5	2.69514906131055\\
3	2.54376345698615\\
3.5	2.40099850937899\\
4	2.26634220889537\\
4.5	2.13931649432423\\
5	2.0194744549139\\
};
\addlegendentry{$\text{Loc-Optimal for SLPR, }\epsilon\text{ = 0.1}$}

\addplot [color=green, dashdotted, line width=2.0pt, mark size=5.0pt, mark=asterisk,mark repeat=5, mark options={solid, green}]
  table[row sep=crcr]{%
-10	708.066996020809\\
-9.5	670.803380358371\\
-9	626.518272471782\\
-8.5	567.823247286797\\
-8	496.679148601909\\
-7.5	438.95952883607\\
-7	385.378831797059\\
-6.5	337.002004624874\\
-6	289.824847453009\\
-5.5	215.618735162803\\
-5	169.297312051676\\
-4.5	133.190327830895\\
-4	106.287200882577\\
-3.5	56.5555617314994\\
-3	38.6805219704389\\
-2.5	24.7784490970968\\
-2	21.7469734467414\\
-1.5	18.3105303482679\\
-1	15.8995069212665\\
-0.5	13.7019367005066\\
0	10.8013391949263\\
0.5	9.56159993329853\\
1	6.93441621877638\\
1.5	6.01110847986857\\
2	4.72916778010591\\
2.5	4.61264458934508\\
3	4.5059369970648\\
3.5	3.7025797563687\\
4	2.41571253664995\\
4.5	2.28009967507048\\
5	2.15219390490567\\
};
\addlegendentry{$\text{Loc-Optimal for SLPR, }\epsilon\text{ = 0.25}$}

\addplot [color=orange, dashed, line width=2.0pt, mark size=5.0pt, mark=square,mark repeat=5, mark options={solid, orange}]
  table[row sep=crcr]{%
-10	596.989343722089\\
-9.5	535.52740309154\\
-9	474.095697820781\\
-8.5	418.410856624483\\
-8	359.387470903667\\
-7.5	308.143616239621\\
-7	249.660149656374\\
-6.5	189.338134326725\\
-6	156.321907882279\\
-5.5	113.414459906984\\
-5	62.487319506391\\
-4.5	43.4176074249023\\
-4	10.1350718600977\\
-3.5	9.41769678099311\\
-3	6.06159197007731\\
-2.5	5.8309748130252\\
-2	4.89396145510955\\
-1.5	4.06533527210695\\
-1	3.83401998105883\\
-0.5	3.6164134976695\\
0	3.41160210011648\\
0.5	3.21875278633488\\
1	3.03710159783775\\
1.5	2.86594526924151\\
2	2.70463422917285\\
2.5	2.55256663698284\\
3	2.40918346108419\\
3.5	2.27396477985465\\
4	2.14642574452603\\
4.5	2.02611378525358\\
5	1.91260568185837\\
};
\addlegendentry{$\text{Loc-Optimal for ISL, }\epsilon\text{ = 0.1}$}

\addplot [color=black, dashdotted, line width=2.0pt, mark size=5.0pt, mark=+,mark repeat=5, mark options={solid, black}]
  table[row sep=crcr]{%
-10	702.360132938641\\
-9.5	665.465810773504\\
-9	624.726281623015\\
-8.5	567.420675731289\\
-8	509.23900033537\\
-7.5	446.575378510655\\
-7	393.351138823548\\
-6.5	345.432280771415\\
-6	296.468759510471\\
-5.5	239.970334935225\\
-5	188.710129395988\\
-4.5	154.059782629041\\
-4	125.636570745676\\
-3.5	67.2879599335805\\
-3	48.1981696595162\\
-2.5	17.6761081154668\\
-2	15.9401577119731\\
-1.5	11.7881446143294\\
-1	8.2410045957893\\
-0.5	6.78809388549359\\
0	5.86146990862546\\
0.5	5.24016667001779\\
1	3.91904983067022\\
1.5	3.77204677759956\\
2	2.85294293363486\\
2.5	2.69207927561633\\
3	2.54048175807976\\
3.5	2.39758170784318\\
4	2.26285141329092\\
4.5	2.13580027260473\\
5	2.01597113897031\\
};
\addlegendentry{$\text{Loc-Optimal for ISL, }\epsilon\text{ = 0.25}$}

\end{axis}

\end{tikzpicture}%
    
    \caption{MML RMSE of range estimation versus SNR for SLPR and ISL maximization with different $\epsilon$ values.}
    \label{fig:mml_snr_all} \vspace{-5mm}
\end{figure}
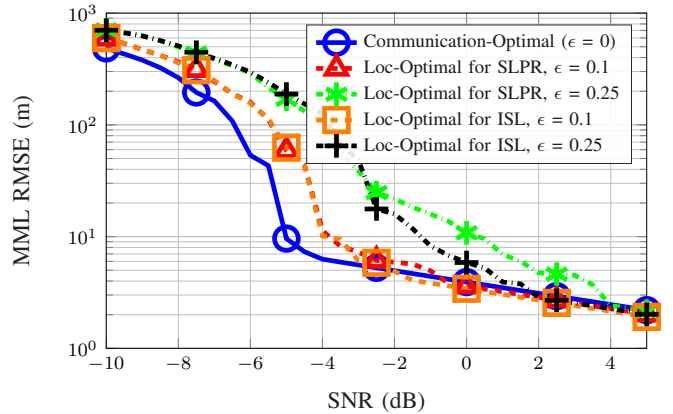

\section{Conclusion}
This manuscript presented a signal-level privacy-preserving framework for OFDM uplinks, where pilot distortions are optimized to degrade ToA estimation accuracy while maintaining reliable communication performance. By formulating the problem as SLPR and ISL maximization under a dissimilarity constraint around the communication-optimal point, the proposed method effectively increases sidelobe power and ToA estimation error with minimal loss in achievable capacity. Simulation results demonstrate that ToA estimation can be selectively degraded without introducing additional delay or altering the transmission protocol, providing a practical and low-complexity approach to privacy-preserving communication.


\appendix[Solution of DoC problem with KKT Points]\label{sec:doc_soln}

In this section, we will solve the optimization problem in \eqref{eq:DoC_simplified} by using the KKT conditions. The Lagrangian of the \eqref{eq:DoC_simplified}
\begin{equation}
    \mathcal{L}(\zz,\lambda)
= \beta\,\zz^{\mathsf{H}}\BB\zz - 2\,\realp{\bb^{\mathsf{H}}\zz}
+ \lambda\big(\norm{\zz-\1}^2-\epsilon P_t\big)\,,
\end{equation}
where $\bb = \AAb\ziter{t}$. Using $\realp{\bb^{\mathsf{H}}\zz}=\tfrac12(\bb^{\mathsf{H}}\zz + \zz^{\mathsf{H}}\bb)$ and $\norm{\zz-\1}^2=\zz^{\mathsf{H}}\zz-\zz^{\mathsf{H}}\1-\1^{\mathsf{H}}\zz+\1^{\mathsf{H}}\1$, we can take the gradient of the Lagrangian, resulting in 
\begin{equation}
    \beta\BB\zz - \bb + \lambda(\zz-\1) = \0\,,
\end{equation}
which leads us to 
\begin{equation} \label{eq:DoC_stationary}
    (\beta\BB+\lambda \mathbf{I})\zz = \bb + \lambda \1 \,.
\end{equation}
Here, if $\zz$ is chosen to be real, $\bb$ needs to be replaced with $\realp{\bb}$ since $\realp{\bb^{\mathsf{H}}\zz}=\realp{\bb^{\mathsf{H}}}\zz$ for real $\zz$. From complementary slackness, we have
    $\lambda\big(\norm{\zz-\1}^2-\epsilon P_t\big)=0$. 
This leads to two cases:
(i) If $\lambda = 0$, the solution is $\zz=(\beta\BB)^\dagger \realp{\bb}$; 
   (ii) If $\lambda>0$, the solution is $\zz(\lambda)=(\beta\BB+\lambda\mathbf{I})^{-1}(\bb+\lambda\1)$, with $\norm{\zz(\lambda)-\1}^2 =\epsilon P_t$ selecting the value of $\lambda$.

Note that until that point, the solution holds for any PSD $\AAb$ and $\BB$. But for our case, we can obtain the closed form result for the second case by using the expansions $\BB = \1 \1^\trp$ and $\norm{\1}^2 = N$. Using the Sherman-Morrison inverse, we have
\begin{equation} \label{eq:left_inv}
    \begin{split}
    (\beta\BB+\lambda \mathbf{I})^{-1} &= (\lambda \mathbf{I} + \beta \1 \1^\trp)^{-1}\\
    &=\frac{1}{\lambda}I-\frac{\beta}{\lambda(\lambda + \beta N)} \1 \1^\trp\,.
    \end{split}
\end{equation}
Plugging \eqref{eq:left_inv} into \eqref{eq:DoC_stationary}, we obtain:
\begin{align}
    &\mathbf{z}(\lambda)
=\frac{1}{\lambda}{\bb}
+\1\left[1-\frac{\beta\,(s+\lambda N)}{\lambda(\lambda+\beta N)}\right] \,,\\
\Rightarrow &\mathbf{z}(\lambda)-\1
=\frac{1}{\lambda}{\bb}
-\1\,\frac{\beta\,(s+\lambda N)}{\lambda(\lambda+\beta N)}\,,
\end{align}
with $s = \1^\trp {\bb}$ As a result, the constraint becomes $ \norm{\zz(\lambda)-\1}^2 = {\norm{{\bb}
-\1\frac{\beta\,(s+\lambda N)}{\lambda+\beta N}}^2}/{\lambda^2 }$. 
Solving this for $\lambda$ to equalize to $\epsilon P_t$, the solution for the second case is obtained.

\balance 
\bibliographystyle{IEEEtran}
\bibliography{IEEEabrv,Sub/PrivacyPreservation}

\end{document}